\documentclass[12pt]{article}

\usepackage{amsmath, setspace}
\usepackage{amssymb, xspace}

\usepackage{graphicx}
\usepackage{color} 

\usepackage{subcaption}
\usepackage{hyperref}
\usepackage{times}
\usepackage[numbers, sort&compress]{natbib} 
\usepackage[table,rgb]{xcolor}
\usepackage{multirow}
\usepackage[misc,geometry]{ifsym}

\usepackage[font=footnotesize]{caption}
\DeclareCaptionLabelSeparator{pipe}{$\; | \;$}
\definecolor{abstarctBlue}{rgb}{0.0706, 0.349, 0.6667} 
\definecolor{abstarctBlue2}{cmyk}{ 0.5143,   0.2857,    0.0286,    0.3714}
\definecolor{Section}{cmyk}{0.2,0.8,0.8,0.3}

\usepackage{soul}
\sodef\an{\fontfamily{phv}\selectfont}{.08em}{1em plus1em}{0.5em plus.1em minus.1em} 
\sodef\ann{\fontfamily{phv}\selectfont}{0.04em}{0.5em plus0.02em}{0.1em plus.1em minus.1em}

\makeatletter
\renewcommand{\@biblabel}[1]{\quad#1.}
\makeatother

\date{}

\usepackage{overpic}
\newcommand*{\hvfont}{\fontfamily{phv}\selectfont}

\newcommand{\fg}{\textcolor{linkcolor}{Figure}~\ref}

\definecolor{citecolor}{rgb}{0.071, 0.36, 0.67}   
\definecolor{linkcolor}{rgb}{0.071, 0.4, 0.67}  
\hypersetup{
colorlinks=true, 
citecolor=citecolor,
linkcolor=linkcolor,
urlcolor=linkcolor
}

\newcommand{\scopems}{SCoPE-MS\xspace	} 
\newcommand{\scope}{SCoPE2\xspace	}

\let\citep=\cite
\let\citet=\cite

\begin{document}
\include{Captions}
\begin{spacing}{1.6}
\noindent {\LARGE \bf 
Single-cell protein analysis by mass-spectrometry 
}
\end{spacing}

\noindent\ann{
Nikolai Slavov$^{1,2,3,}$\textsuperscript{\Letter }
} \\

{\small 
\noindent 
$^{1}$Department of Bioengineering, Northeastern University, Boston, MA 02115, USA\\
$^{2}$Barnett Institute, Northeastern University, Boston, MA 02115, USA\\
$^{3}$Department of Biology, Northeastern University, Boston, MA 02115, USA\\
}
{\scriptsize \Letter} Correspondence: \href{mailto:nslavov@alum.mit.edu}{\an{\small nslavov@alum.mit.edu}} or \href{mailto:nslavov@northeastern.edu}{\an{\small nslavov@northeastern.edu}} \\
\thispagestyle{empty}
\vspace{1cm}

\vspace{10mm}
\noindent{\bf
Human physiology and pathology arise from the coordinated interactions of diverse single cells. However, analyzing single cells has been limited by the low sensitivity and throughput of analytical methods. DNA sequencing has recently made such analysis feasible for nucleic acids, but single-cell protein analysis remains limited. Mass-spectrometry is the most powerful method for protein analysis, but its application to single cells faces three major challenges: Efficiently delivering proteins/peptides to MS detectors, identifying their sequences, and scaling the analysis to many thousands of single cells. These challenges have motivated corresponding solutions, including SCoPE-design multiplexing and clean, automated, and miniaturized sample preparation. Synergistically applied, these solutions enable quantifying thousands of proteins across many single cells and establish a solid foundation for further advances. Building upon this foundation, the SCoPE concept will enable analyzing subcellular organelles and post-translational modifications while increases in multiplexing capabilities will increase the throughput and decrease cost.
}
\vspace{10mm}

\begin{spacing}{1.08}
\section*{Introduction}
Mass spectrometry (MS) allows quantitative protein analysis at large scale \citep{cravatt2007biological, Aebersold_Mann_2016}. Yet when applied to populations of cells, such as those comprising tissues, MS measurements usually average out the differences between the diverse cell types comprising the tissues. These average protein abundances in a tissue cannot be used to reliably infer protein levels in each of the cells comprising the tissue. This problem is well recognized and has motivated the development of numerous approaches for reducing the confounding effects of averaging across cell types \citep{Arjun_2016_Noise, Levy_Review_2018, Specht_Perspective_2018, Nature2019voices}.  

Averaging artifacts may be partially mitigated by first isolating cells from each type based on molecular markers and then separately analyzing groups of cells from each cell type \citep{di2011highly, myers2019streamlined}. This simple approach assumes that (i) we have good molecular markers for each cell type and (ii) that the cells isolated based on a set of markers are not functionally diverse. Both assumptions are frequently violated. First, the molecular markers may not be known, maybe difficult to measure (e.g., because of lack of good antibodies), or the markers needed to separate multiple subpopulations may be too numerous to be feasible to separate all subpopulations. Second, bulk analysis of the isolated cells cannot test their homogeneity. We may assume that the isolated cells are homogeneous, but this assumption cannot be evaluated and falsified by bulk analysis of the isolated cells. Consider, for example, profiling immune cells. B and T lymphocytes can be isolated from blood samples using well-defined markers (e.g., CD3 for T-cells and CD19 for B-cells), but heterogeneity within each isolated subpopulation will be obscured by measuring the average RNA and protein abundances in the subpopulations \citep{myers2019streamlined, wagner2020lineage}. The heterogeneity of the isolated cells only becomes apparent through single-cell analysis. Indeed, single-cell analyses have recently demonstrated the existence of multiple states within T-cell sub-populations, although these states rarely have well-defined markers to enable efficient FACS isolation and downstream bulk analysis \citep{Nature2019voices, savas2018single}. Bulk analysis of isolated cells is particularly limited when cellular states do not fall into discrete subpopulations but rather define continuous cycles \citep{Slavov_batch_ymc, Slavov_emc} or gradients, as found to be the case with macrophages differentiated in the absence of polarizing cytokines \citep{scope2}.

These limitations of bulk analysis can be relaxed by performing single-cell analysis. Indeed, single-cell analysis by RNA sequencing has began to trace cell lineages and to find physiologically relevant differences within cells that were considered homogeneous \citep{savas2018single, Semrau_Nemes, wagner2020lineage}. Despite this exciting progress, RNA levels are insufficient to characterize and understand biological functions arising from post-transcriptional regulation, which is wide spread in human tissues \citep{Franks2016PTR}. RNA measurements do not reflect protein degradation,  protein interactions (such as complex formation), post-translational modifications and re-localization (such as transcription factors localizing to the nucleus or mTOR localizing to the lysosomal surface) \citep{Slavov2020Science}. These post-transcriptional mechanisms are better characterized by direct measurements of proteins in single cells. 

For the last two decades, such single-cell protein measurements have relied on antibody-based methods \citep{Levy_Review_2018}. These methods have made major contributions, but they remain rather limited by antibody availability and specificity and by the number of proteins that can be analyzed simultaneously \citep{Levy_Review_2018, Slavov2020Science}. These limitations can be overcome by emerging mass-spectrometry (MS) methods. Below we review the challenges for MS methods and approaches that have provided productive solutions in the last few years. While  single-cell protein analysis is the focus, by many of the challenges and solutions are applicable to other types of single-cell MS analysis, such as single-cell metabolite analysis \citep{Semrau_Nemes}.


\section*{Challenges to single-cell mass-spectrometry analysis}
Protein analysis by MS generally includes sample preparation, peptide/protein separation (usually by liquid chromatography or capillary electrophoresis), ionization, and tandem MS analysis. These steps have been reviewed in-depth by Ref. \citep{Aebersold_Mann_2016, cravatt2007biological, sinitcyn2018computational}, and Ref. \citep{sinitcyn2018computational} also provides an excellent description of data interpretation and downstream analysis. Each of these steps brings challenges for analyzing very small samples, such as single-cell proteome.

Most proteins are present at thousands of copies per cell while mass-spectrometry detectors can detect and quantify hundreds of ion copies per MS scan, even a single ion copy \citep{worner2020resolving, Neil2020SingleIons}.  Thus, the sensitivity of detectors is generally not the major limitation. Rather, the major challenges are (i) delivering proteins to the MS detectors, (ii) identifying the sequence of peptide or protein ions, and (iii) analyzing proteins from many thousands of single cells at affordable cost, see \fg{honeycomb}. These challenges have shaped the approaches to single-cell MS analysis for the last three decades. This review  summarizes successful strategies for overcoming the challenges, starting with a short overview of early efforts, and focusing on recent advances that have established the foundation for quantitative analysis of proteins at single-cell resolution.

\subsection*{Early approaches to ultrasensitive mass-spectrometry analysis}
The first challenge, delivering proteins from small samples to MS detectors, was initially approached by employing Matrix-Assisted Laser Desorption/Ionization (MALDI) as a means to ionizing peptides and proteins. MALDI allows to ionize proteins with minimal sample handling and surface exposure. Thus, MALDI helps to minimize losses and to deliver ions to the MS detectors, usually time-of-flight (TOF) detectors. Using MALDI-TOF approaches, multiple groups were able to detect proteins from single cells in the 1990s \citep{veelen_direct_1993,li_situ_1999}; Ref \citep{boggio2011recent} offers detailed review. However, MALDI approaches usually do not separate peptides in time and only a few of the detected ions can be sequenced. Thus, single-cell protein analysis by MALDI has been limited by the second challenge, determining the amino acid sequence. Furthermore, the variability in MALDI ionization undermines quantification accuracy. 

The other major approach to ionizing proteins and peptides, electrospray ionization (ESI), is more amenable to sequencing detected ions since it is more readily coupled to peptide separation methods \citep{Aebersold_Mann_2016}. However, sample handling and separation prior to ESI may result in more sample losses. Nonetheless, ESI has also been used since 1990s for analyzing abundant proteins in small samples. Indeed,  hemoglobin was detected in samples comprised of a few erythrocytes \citep{hofstadler1995capillary} or a single erythrocyte \citep{hofstadler1996analysis}.     
Yet these methods did not generalize to analyzing many proteins in typical mammalian cells: Hemoglobin is present at 300 million copies per erythrocyte, about 6,000 fold more abundant that the median abundance protein in a typical mammalian cell, such as a fibroblast \citep{milo2010bionumbers}. 

An important early advance in ultrasensitive MS analysis via ESI was the use of capillary electrophoresis (CE) \citep{hofstadler1995capillary, valaskovic1996attomole}. CE allows using small sample volumes, and thus may enhance sample delivery to MS detectors (challenge one in \fg{honeycomb}). Therefore, CE has been an effective means for  separation and sensitive analysis of both proteins and metabolites in very small samples. As discussed below, CE-MS analysis continues to drive progress in single-cell proteomics \citep{lombard2016single, belov2017analysis,  lombard2019microsampling}.  
\newpage

\section*{Synergistic approaches advancing single-cell proteomics}
Multiple recent advances have made major contributions to overcoming the challenges to singe-cell MS analysis, namely to improve the delivery of proteins, to enhance sequence determination, and to increase throughput, \fg{honeycomb}. These advances combine synergistically to enable quantitative analysis of thousands of proteins across many single cells \citep{SCP_review_2019, Ben_bioAnalysis, Slavov2020Science}. To systematically review these advances, they are grouped into the categories displayed in \fg{honeycomb} and discussed below.\\[0.5em]

\begin{figure}[h!]
    \centering
    \includegraphics[width = 0.82\textwidth]{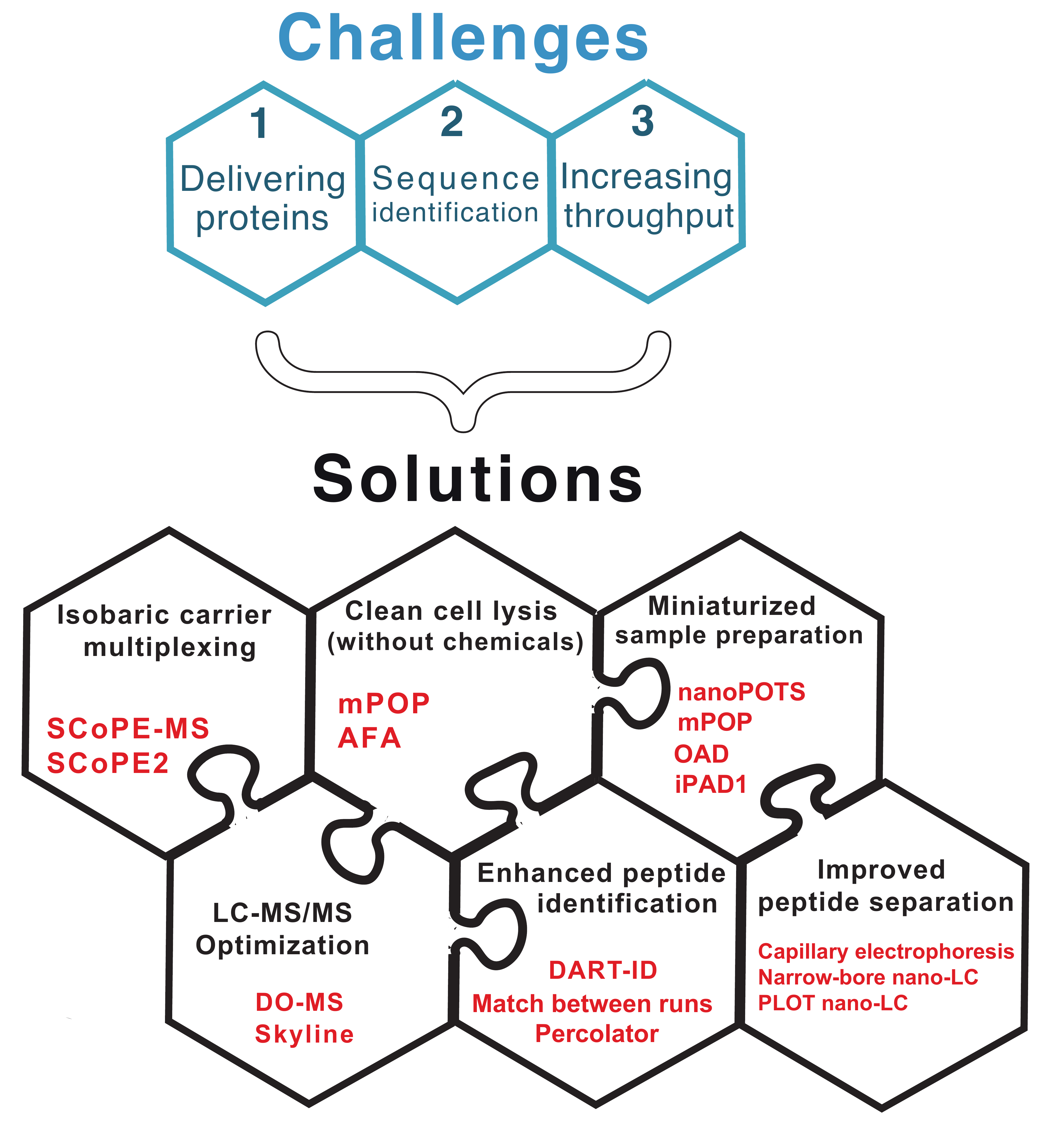}
    \caption{{\bf \hvfont \scriptsize Major challenges to single-cell protein analysis by mass-spectrometry and their solutions} \hvfont \scriptsize Single-cell proteomics by mass-spectrometry faces three main challenges: (i) Delivering enough copy number of ions to MS detectors to afford accurate quantification; (ii) Reliable amino acid sequence determination of quantified ions; (iii) Scaling the analysis to many thousands of single cells at affordable cost. Each of these challenges is addressed by one or more solutions. These solutions are highly synergistic and thus depicted as interlocking puzzle pieces with a few prominent examples from each category. }
    \label{honeycomb}
\end{figure}

\subsection*{Multiplexing with isobaric carrier} 
In response to the first and the second challenge shown in \fg{honeycomb}, we developed Single Cell ProtEomics by Mass Spectrometry (SCoPE-MS) \citep{scopems2017, scopems2018, Specht_Perspective_2018}. \scopems introduced an isobarically-labeled carrier concept, which is abbreviated below to \emph{isobaric carrier}.  Carrier proteins and peptides have long been used for passivating surfaces and reducing adsorbent losses. The isobaric carrier approach introduced by  \scopems is different in employing a carrier that is labeled by isobaric mass tags and that is used in the MS analysis. Specifically, the SCoPE carrier approach employs tandem mass tags \citep{thompson_tandem_2003} to label lowly abundant samples of interest (e.g., single-cell proteomes) and a carrier sample (e.g., the proteome of 100 cells), and then combines all labeled samples to be analyzed together by liquid chromatography tandem mass-spectrometry, \fg{scope2}. The use of TMT ensures that copies of a given peptide sequence from the single-cell and bulk samples all have the same mass-to-charge ratio during survey scans, so that they are isolated together for fragmentation and MS2 analysis. During the fragmentation, the precursor ions  generating sample-specific reporter ions whose abundances allow for relative quantification \citep{thompson_tandem_2003, sinitcyn2018computational}.  Thus the isobaric carrier approach helps to (i) mitigate losses from the small samples (since adsorption losses will disproportionately affect the carrier proteome), (ii) increase peptide sequence identification (since the carrier proteome will provide peptide fragments), and (iii) increase throughput (since multiplex labeling with TMT allows simultaneous analysis of multiple samples). Therefore, the isobaric carrier design introduced with SCoPE-MS mitigates the three major challenges displayed in \fg{honeycomb} \citep{scopems2017, scopems2018, SCP_review_2019, Ben_bioAnalysis}. \\[0.5em]  

\begin{figure}[h!]
    \centering
    \includegraphics[width = 0.9\textwidth]{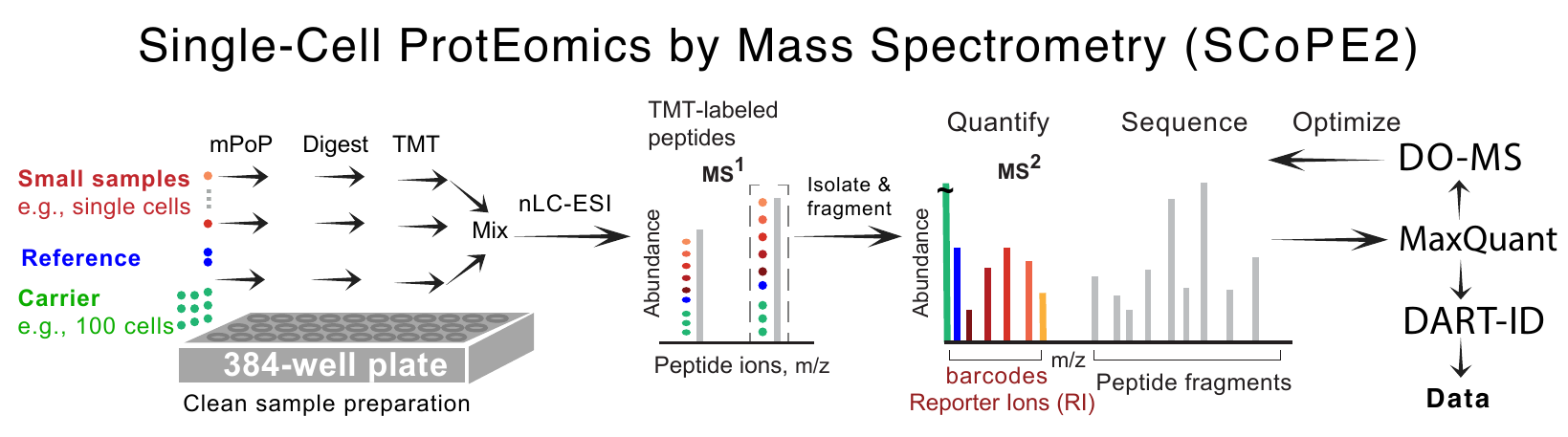}
    \caption{{\bf \hvfont \scriptsize Conceptual work flow of automated sample preparation, the isobaric carrier design, enhanced peptide sequence identification, and LC-MS/MS optimization as implemented by \scope}  \hvfont \scriptsize \scope provides solutions for all challenges from \fg{honeycomb}: (i) Protein delivery for MS analysis is facilitated by the clean, automated and miniaturized lysis by mPOP \citep{mPOP_2018}, and by the isobaric carrier design, which combines isobarically-labeled peptides from single-cell and from carrier samples \citep{scopems2017,scopems2018,scope2}. (ii) Peptide sequence identification is enhanced by the carrier peptides contributing fragment ions to the MS2 spectra and by DART-ID \citep{dartID_PLoS}.  (iii) Analysing many cells is facilitated by fully automated sample preparation and analysis, and by the isobaric carrier design multiplexing with TMT pro \citep{mPOP_2018, scope2, Specht_Perspective_2018, Levy_Review_2018}.}
    \label{scope2}
\end{figure} 

The simplicity and effectiveness of the isobaric carrier design have stimulated its initial adoption. Since its introduction in January 2017 \citep{scopems2017}, multiple groups have used the approach for goals including the detection of rare proteoforms \citep{Variant_detection_JPR}, phosphorylation \citep{yi2019boosting, BOOST_tyrosine_phosphoproteome}, translation measurements \citep{klann2020functional} and single-cell protein analysis with \scopems \citep{scopems2018, Schoof_SCeptre, PNNL_2019, iBASIL}; reveiwed by Ref. \citep{SCP_review_2019, Ben_bioAnalysis}. A related approach (TMTcalibrator\textsuperscript{TM}) mixed TMT-labeled samples from plasma and cell lines to identify markers of microglia activation \citep{russell2017combined}.  Some authors  used ``booster'' as a  synonym of ``isobaric carrier''. Regardless of the term used, the concept is the same: The carrier sample helps to reduce losses from the single cells and to enhance peptide sequence identification. However, the carrier sample does not amplify or boost the single-cell reporter ion intensities. Therefore, it is essential to ensure that the MS analysis delivers a sufficient number of ion copies from the single-cell peptides to support reliable quantification \citep{Specht_Perspective_2018, scope2}.  

The SCoPE design can afford accurate quantification of protein changes across single cells (i.e, relative quantification) based on the reporter ions shown in \fg{scope2} \citep{scopems2018, scope2, PNNL_2019, Schoof_SCeptre, iBASIL, SCP_review_2019, Ben_bioAnalysis}. However, the accuracy of comparing the abundances of different proteins is lower. This weakness can be overcome by three different approaches. First, principled models, such as HI\emph{quant}, can estimate protein stoichiometries (i.e., compare abundances of different proteins) using only relative quantification \citep{HIquant-MCP}. Second, spiked-in standards with known absolute quantification can be used to estimate absolute protein abundances, i.e., number of protein copies per cell. Then, these absolute estimates can enable comparison of the abundances of different proteins and proteoforms  \citep{HIquant-MCP}. Third, absolute protein abundances can be estimated from bulk analysis of the heterogeneous cell populations (e.g., the carrier samples), and then these estimates of absolute abundances can be apportioned to the single-cell samples in correspondence to the relative protein levels measured in single cells.

\subsection*{Clean and automated sample preparation}
Bulk samples are typically prepared for MS analysis by lysing the cells in buffers containing detergents and other chemicals that if not cleaned may affect adversely enzymatic processes (such as protease digestion) and MS analysis. Thus, for optimal results such chemicals should be removed. While the losses incurred by these clean-up steps are usually acceptable with bulk samples, they are less tolerable with very small samples, such as single mammalian cells. Furthermore, these cleanup steps complicate automation, which is an essential aspect of maximizing the number of single cells that can be analyzed while minimizing the cost and batch effects \citep{Specht_Perspective_2018}.  

Several reports have demonstrated that adaptive focused acoustics (AFA) can extract proteins for MS analysis \citep{Ivanov2015RareCells, dhabaria2015high}, and AFA was a natural method to use when we started developing single-cell proteomics methods in the fall of 2015 \citep{scopems2017, scopems2018}.  While AFA allowed lysing mammalian cells without using MS incompatible chemicals \citep{Ivanov2015RareCells, dhabaria2015high}, it required a sample volume of $10\mu l$ per cells, and we sought to develop methods with reduced volumes, as elaborated below. Furthermore, the automation of AFA required expensive equipment. To overcome these weaknesses, we developed and validated a second generation method, Minimal ProteOmic sample Preparation (mPOP) \citep{mPOP_2018, nature_highlight}. mPOP uses a freeze-heat cycle to efficiently deliver proteins to MS analysis and afford reliable quantification of proteins in single cells \citep{scope2}. Importantly, mPOP allowed us to reduce sample volumes 10-fold (to $1 \mu l$ / cell) and to completely automate sample preparation with inexpensive equipment \citep{mPOP_2018, scope2}

\subsection*{Miniaturized cell lysis and sample preparation}
Decreasing the volume of sample preparation reduces the amount of reagents that have to be added (e.g., trypsin, buffers, tandem mass tags). It also reduces the surface areas contacting the sample and thus the potential for losses from proteins adhering to surfaces during sample preparation.  These considerations led to the development of methods for small volume sample preparation. 

As discussed above, mPOP allows sample preparation in standard multi-well plates in volume of $1 \mu l$ / cell,  and further reductions in volume are still desirable \citep{Specht_Perspective_2018, nature_highlight}. Indeed, several groups have developed methods that afford lower volumes, down to hundreds of nanoliters. These methods include nanodroplet processing in one pot
for trace samples (nanoPOTS) \citep{zhu2018nanodroplet}, oil-air-droplets (OAD) \citep{li2018nanoliter}, and on column cell lysis by iPAD1 \citep{shao2018integrated}, and they have allowed identifying from a few dozen to hundreds of proteins from label-free analysis of individual cells \citep{li2018nanoliter, shao2018integrated, zhu2018proteomic, cong2020improved}. nanoPOTS has also been used to prepare single cells for \scopems experiments (i.e., with isobarically-labeled carriers) that identified over a thousand proteins  \citep{PNNL_2019, iBASIL}. 
Since automated loading of hundreds of nanoliters on chromatographic columns is challenging, cells prepared by these methods are usually loaded manually. Thus, taking full advantage of  miniaturized cell lysis and sample preparation requires further miniaturization of all other steps of the analysis, including efficient and automated loading of small samples on chromatographic columns.   

As discussed above, miniaturized sample preparation allows reducing adhesion related sample losses, which in turn enables analyzing individual cells by ultrasensitive label-free LC-MS/MS analysis \citep{li2018nanoliter, shao2018integrated, zhu2018proteomic, cong2020improved}. Such label-free analysis simplifies sample preparation since it does not require labeling steps. Furthermore, it obviates the use of chemical labels, which may contribute chemical impurities that interfere with sample separation or MS analysis. Label-free MS analysis  already affords detecting hundreds of protein groups in single  Hela cells, and improvements in peptide separation, ionization and in MS instrumentation are going to further increase the number of proteins that can be detected and quantified by label-free methods \citep{cong2020improved}. Currently, these ultrasensitive LC-MS/MS workflows tend to have manual  steps, such as transferring samples to chromatographic columns. Automating manual steps is likely to increase the throughput of single-cell label-free methods. Still, methods that use chemical labeling for multiplexing are likely to afford higher throughput and lower cost by enabling the simultaneous analysis of multiple samples \citep{lombard2016single}. Indeed, approaches using the isobaric carrier concept for multiplexing have analyzed over a thousand single cells while label-free approaches have analyzed only a few single cells.

\subsection*{Optimizing parameters for data acquisition by LC-MS/MS}
Every LC-MS/MS experiment depends upon numerous parameters whose optimization can substantially increase the number of identified peptides and the accuracy of their quantification \citep{bittremieux_computational_2017, sinitcyn2018computational}. Such optimization is particularly important for the analysis of small samples, such as single-cell proteomes, since even low levels of contaminants or reductions in ion delivery may substantially undermine data quality \citep{sinitcyn2018computational, Specht_Perspective_2018, nature_highlight}.   

To facilitate benchmarking and optimization LC-MS/MS experiments, we developed Data-driven Optimization of Mass-Spectrometry (DO-MS) \citep{doMS2019}. DO-MS aims to specifically diagnose problems in LC-MS/MS analysis by interactively visualizing data from all levels, from the peptide separation and the survey scans to ion isolation for MS2 analysis and matching spectra to sequences, and thus has become an integral part of \scope, \fg{scope2}. One aspect that has benefited significantly by the DO-MS application is improving apex targeting. Specifically, DO-MS visualizes time offsets from the elution peak of a peptide (its apex) and the time when the peptide is sampled for MS2 analysis. These data allowed us to rationally correct for systematic biases, such sampling elution peaks too early \citep{doMS2019}. While it is not possible to deterministically ensure apex sampling for every peptide, a higher proportion of peptides can be sampled at or near their elution apices by choosing optimal LC-MS/MS parameters. Via data visualization platforms like DO-MS, LC-MS/MS parameters, such as the maximum number of MS2 scans triggered per MS1 scan, can be iteratively tuned until system suitability is optimal for single-cell samples. Many tools for diagnosis and optimizing LC-MS/MS have not be designed specifically for single-cell proteomics but can nonetheless be very useful; such methods are reviewed by Bittremieux, et al \citep{bittremieux_computational_2017}.

\subsection*{Enhancing peptide sequence identification}
The abundance of a peptide may be estimated based on a single mass/charge peak that originates from the peptide. However, determining its sequence generally requires multiple fragment ions, some of which are produced with low efficiency and may not be detectable in lowly abundant sample \citep{eng1994approach}. Thus, peptides whose abundance is quantified may be challenging to identify. Sequence identification is particularly challenging with MS methods that allow for limited peptide fragmentation, such as MALDI-TOF \citep{boggio2011recent}.    

To alleviate these difficulties in determining peptide sequence, it is desirable to use all features informative for the sequence as implemented by the  Percolator \citep{percolator2009improvements} and recently applied to \scopems datasets \citep{fondrie2020machine}. The retention time (RT) and ion mobility of a peptide can be very informative features for its sequence, and RT has been used by many methods, including for disambiguating mixed spectra by CharmeRT \citep{dorfer2018charmert} and for increasing peptide identifications by Skyline ion matching \citep{maclean2010skyline}, and MaxQuant match-between-runs \citep{tyanova2016maxquant, yu2020isobaric}. Yet, these methods do not necessarily estimate the false discovery rate (FDR) of peptide sequences determined based on the retention time. To fill in this gap, we developed a principled Bayesian framework for incorporating retention time information in determining peptide sequences, Data-driven Alignment of Retention Times for IDentification (DART-ID) \citep{dartID_PLoS}. DART-ID can be applied to most MS datasets, and it is particularly powerful for single-cell proteomics. It can increase the number of confidently identified peptides by 50\% at 1\% FDR and contributes significantly to the \scope framework as shown in \fg{scope2} \citep{dartID_PLoS, scope2}

\subsection*{Improving peptide separation and ionization}
The importance of high performance  peptide separation and ionization has been an integral part of developing ultrasensitive MS analysis for decades \citep{valaskovic1996attomole, altelaar2012trends}. Sharp elution peaks and low flow rates help maximize the delivery of proteins to the MS detectors \citep{Specht_Perspective_2018, cong2020improved}. These principles have been implemented both by capillary electrophortesis \citep{hofstadler1995capillary, valaskovic1996attomole, lombard2016single, belov2017analysis, lombard2019microsampling},  by highly sensitive multidimensional chromatographic strategies \citep{di2011highly, cifani2017high}, and by liquid chromatography using  monolithic nanocapillary columns, PLOT columns, small bore columns and low flow rates \citep{ivanov2003low, Ivanov2015RareCells, cong2020improved}. 

Shorter chromatographic gradients and electropherograms help maximize the number of single-cells samples analyzed per unit time. Indeed, reducing the gradient length from 180 min for \scopems to 60 min for \scope helped increased throughput without concomitant decrease in the number of quantified proteins \citep{scope2}.  Improved separation is a very important aspect of ultrasensitive MS analysis has been extensively reviewed, e.g. by Ref. \citep{altelaar2012trends, shishkova2016now}.

\section*{Future Developments}
Recently, the power of single-cell protein analysis by MS has increased by orders of magnitude \citep{Specht_Perspective_2018, scope2}. This growth marks the beginning of a new phase whose growth will likely continue and even accelerate. This future growth will build upon and extend the advances outlined in \fg{honeycomb}. Below are highlight some promising directions, both for extending approaches that are already fruitfully applied to single-cell analysis and for introducing new ones.

\subsection*{Extending the SCoPE design}
The ability of the isobarically labeled carrier proteins to influence the ions selected for MS2 analysis can be exploited to target the analysis of protein modifications, sub-cellular structures, or any defined group of proteins under investigation. As previously suggested \citep{Slavov2020Science}, if the carrier channel contains post-translationally modified peptides (e.g., phosphorylated peptides enriched by immobilized metal affinity chromatography), the most abundant ions detected in survey scans will correspond to the phosphorylated peptides from the carrier channel, and thus they will be selected for MS2 analysis, quantification and identification. Similarly, if the carrier contains mitochondria, mitochondrial proteins will be selected for MS2 analysis. Of course, selection for MS2 analysis does not guarantee clean spectra and quantification in the single-cell samples. Achieving reliable single-cell quantification requires reducing coisolation effects (e.g., good apex targeting or using complement ions as discussed below) and delivering sufficient ion copy numbers from each single cells. These aspects must be rigorously benchmarked before one can confidently extend the SCoPE concepts more broadly to quantifying post-translational modifications and and sub-cellular structures.   

Extending the \scope framework to large scale targeted analysis can increase the sensitivity (e.g., by increasing ion accumulation times), the reliability (e.g., by sampling more ion copies per peptide), and the reproducibility (e.g., by consistently sampling the same precursor ions)  of single-cell MS analysis \citep{Specht_Perspective_2018}.  Such targeted analysis may afford consistent sampling and quantification of thousands of proteins, and thus reduce missing data, which is a common problem in high-throughput single cell analysis \citep{Nature2019voices, savas2018single, scope2}.

\subsection*{Increasing multiplexing}
Sample throughput scales with the number of available tandem mass tags (chemical barcodes), and we have already demonstrated 50 \% increase in throughput due to increased multiplexing \citep{scope2}. With \scopems, we used 10-plex TMT labels and could analyze only 8 single-cell samples per LC-MS/MS run. With \scope, we used 16-plex TMT pro labels and could analyze 12 single-cell samples per LC-MS/MS run \citep{scope2}. Extrapolating this trend to a hundred isobaric labels, we expect to analyze the proteomes of about 2,400 single cells in 24 hours of continuous instrument operation. Additionally, as the number of single cells analyzed per run grows, the necessity of the carrier channel diminishes, especially its role in providing peptide fragments for sequence determination. This increased multiplexing is likely to proportional decrease the cost per single cells since at the moment the cost is dominated by the cost of LC-MS/MS time \citep{scope2}. The decreased cost and increased throughput will provide the large-scale data required for many promising biomedical applications \citep{Slavov2020Science}.

\subsection*{Limits of multiplexing}
One fundamental limit on the number of labeled samples, $N$, is set by the capacity of the MS analyzer, $C_{max}$: On average the MS analyzer can sample about $C_{max} / N$ ion copies per sample, and thus for very large $N$ the number of sampled ion copies will not be enough to support reliable quantification. Measurements will be dominated by counting noise. 
For the current orbitrap detectors, $C_{max} \approx 10^6$ ions, and thus $N = 1,000$ will result in sampling on average up to $1,000$ ion copies from each peptide per sample. The sampling error then can be estimated from the Poisson distribution as \emph{standard deviation / mean} to be  $\sqrt{1,000}/1,000 = 3 \%$. Less abundant peptides will have large sampling error while more abundant peptides smaller. 

This limit of multiplexing deserves special consideration in the context of experimental designs including isobaric carriers (i.e., \scopems and \scope) since the carrier sample represents a substantial fraction of analyzed ions, and thus it might disproportionately fill in the orbitrap and leave insufficient space for single-cell peptides. For a carrier sample that is about 200 times larger than the small samples (i.e., single cells), the mass-analyzer can sample on average up to $C_{max} / (200 + N)$ ion copies. With TMT pro, this corresponds to about 47,000 ion copies per peptide from a single cell and a sampling error of about $1.5\%$. In practice, we rarely reach this limit because even the peptide quantity pooled across the carrier and the single cells is too low to reach this limit. As a result, the number of ions accumulated for MS2 analysis is much smaller than the capacity of the orbitrap \citep{scope2}. Indeed, we have observed that the copy number of ions sampled from a single-cell is limited not by the carrier amount but by the efficiency of delivering peptide ions to MS2 scans \citep{Specht_Perspective_2018, scope2}.

\subsection*{DIA analysis of multiplexed single-cell samples}
Modern single-cell MS analysis has focused on the sequential analysis of individual peptide precursors. A well-recognized weakness of this analysis is that relatively few peptides can be analyzed per MS run, especially when the analysis time per peptides is long. In the case of single-cell MS analysis, the analysis time is long because of the need to accumulate enough ions for reliable quantification and sequence identification \citep{scopems2017, Specht_Perspective_2018, scope2}.

An alternative to such sequential analysis was introduced by Yates and colleagues in 2004 \citep{venable_automated_2004} and further developed by Aebersold and colleagues \citep{gillet_targeted_2012}. This alternative is known as data independent analysis (DIA). DIA simultaneously isolates and fragments multiple peptides in parallel. This parallel analysis allows to increase the number and reproducibility of the analyzed peptides. However, if DIA is applied to samples labeled with isobaric mass tags, e.g., the \scope design, the parallel isolation of multiple peptides means that the abundance of the detected reporter ions will reflect the cumulative abundance of all isolated peptides, making it challenging to quantify individual peptides. Thus, DIA has not yet been extended to the analysis of TMT labeled samples.

Extending DIA analysis of \scope samples is both very challenging and promising. One approach would require to quantify single-cell peptides based on the TMT fragments remaining bound to the peptide fragments, known as a mass balancers or complementary ions. These complementary ions have allowed quantifying bulk samples by data dependent methods \citep{sonnett_accurate_2018}, and have not yet been employed in DIA analysis. Such employment will be very challenging but seems feasible. A primary challenge, especially for single-cell analysis, will be sampling enough from these lowly abundant ions to achieve reliable quantification.  A second approach would require to quantify the single-cell reporter ions of each peptide across a large number of MS2 scans so that the superposition of reporter ion intensities can be fit into a linear model and deconvoluted. This approach is also very challenging for single-cell analysis since the long ion accumulation times make it harder to acquire many MS2 scans across the elution profiles.  \\

\subsection*{Advances in MS instrumentation}
Advances in MS instrumentation can also play important role in solving the challenges of  single-cell MS analysis \fg{honeycomb}. As discussed above, instrumentation that allows for automated and reliable loading of very small samples to CE and LC columns is essential for automating high-performance separation, and enhancing ionization by minimizing flow rates. Similarly, improvements in the efficiency of ionization and ion accumulation can improve ion delivery to the MS detectors \citep{marginean2010achieving, kelly2010ion, Specht_Perspective_2018}. Trapped Ion mobility, as implemented by the timsTOF, can both enable parallel accumulation of ions and provide an additional feature (i.e., ion mobility) for peptide sequence identification \citep{Specht_Perspective_2018, ridgeway2019trends}.  \\

\noindent 
The recent progress and futures prospects outlined here promise to bring to single-cell analysis the power and versatility of MS methods that so far have been limited to bulk samples. The resulting single-cell MS methods will far exceed the power of antibody-based methods that so far have dominated single-cell protein analysis \citep{Levy_Review_2018}. If we succeed in making single-cell MS methods robust, inexpensive and widely accessible \citep{Slavov2020Science}, they will become a major enabling factor in identifying molecular mechanisms that underlie health and disease.

\end{spacing}
\vspace{10mm}

\bigskip
\noindent {\bf Acknowledgments:} I thank professors B.L.~Karger and A.R.~Ivanov, as well as  R.~G.~Huffman and H.~Specht  for discussions and constructive comments. This work was funded by a New Innovator Award from the NIGMS from the National Institutes of Health to N.S. under Award Number DP2GM123497.  Funding bodies had no role in data collection, analysis, and interpretation. \\



\subsection*{References and recommended reading}
Papers of particular interest, published within the period of review, have been highlighted as:

\begin{itemize}
    \item [$\bullet$] of special interest
    \item [$\bullet$ $\bullet$] of outstanding interest
\end{itemize}

\bibliographystyle{/Users/nslavov/GoogleDrv/B/texmf/bst/plos-natbib} 
\bibliography{SCoPE,Review,NSrefs}

\end{document}